\pdfoutput=1
\documentclass[10pt,prl,aps,twocolumn]{revtex4-2}
\usepackage{amsmath,amssymb,mathtools,simpler-wick}
\usepackage{bm}
\usepackage{graphicx} 
\usepackage[dvipsnames,svgnames,x11names]{xcolor}
\usepackage{comment}
\usepackage[colorlinks=true,citecolor=blue,urlcolor=blue,linkcolor=blue,pdfstartview=FitH,bookmarksopen]{hyperref}
\DeclareMathOperator{\sgn}{sgn}
\newcommand\+\dagger
\renewcommand\d\partial
\newcommand\x{\mathbf{x}}

\begin{document}

\title{Lifshitz Phase of Quantum Gravity:\texorpdfstring{\\}{ } The Infrared Fixed Point of the (2+1)D Unimodular Ho\v rava--Chern--Simons Model}
\author{Maxim Zelenko and Dam Thanh Son}
\affiliation{Leinweber Institute for Theoretical Physics, University of Chicago, Chicago, Illinois 60637, USA}

\date{October 2026}

\begin{abstract}
We consider a nonrelativistic gravity theory in (2+1) dimensions with $w_\infty$ gauge symmetry that we call unimodular Ho\v rava--Chern--Simons gravity.  The theory has three dimensionless couplings, one of which is the Chern--Simons level $\kappa$.  At large $|\kappa|$, the theory flows from an asymptotically free UV fixed point to an IR fixed point with anisotropic scale invariance characterized by a dynamical critical exponent $z=2-\tfrac32|\kappa|^{-1}$.  We speculate that unimodular Ho\v rava--Chern--Simons gravity undergoes a phase transition from this scale-invariant phase to a confining phase when $\kappa$ drops below a critical value.

\end{abstract}

\maketitle

\emph{Introduction.}---One of the biggest challenges of modern physics is to reconcile quantum mechanics with general relativity.  The difficulty comes from the well-known nonrenormalizability of Einstein gravity---the coupling constant has negative mass dimension, so the theory is not under control in the ultraviolet (UV) \cite{Goroff:1985th}.  This was the main motivation for Ho\v rava's proposal of a new type of gravitational theory \cite{Horava:2009uw}.  Ho\v rava gravity does not have Lorentz invariance; instead, at the classical level, there is an anisotropic scale invariance with the dynamical critical exponent $z=d$ in $(d+1)$ dimensions.  The theory is power-counting renormalizable, and the one-loop running of the coupling constants has been studied in the so-called  projectable version, where the number of coupling constants is greatly reduced because the lapse field depends only on time \cite{Sibiryakov,Barvinsky:2021ubv}.  It was found that the theory is asymptotically free in (2+1) dimensions \cite{Sibiryakov}.  The structure of fixed points and RG flow is more complicated in (3+1)D projectable Ho\v rava gravity \cite{Barvinsky:2024svc}. 

In this Letter, we consider a gravitational theory that gives rise to a scale-invariant phase [i.e., an infrared (IR) stable fixed point] in (2+1) dimensions.  We call that theory \emph{unimodular Ho\v rava--Chern--Simons} (UHCS) gravity: as described below, it is a Ho\v rava-type gravity, modified by the unimodular constraint on the metric and the corresponding restriction on the gauge symmetry, the removal of the lapse field, and a Chern--Simons (CS) term with level $\kappa\neq0$ in the action. Like pure Ho\v rava gravity, the theory is power-counting renormalizable. The renormalization group (RG) flow is from an asymptotically free UV fixed point with $z=2$ to a scale-invariant IR fixed point with a nontrivial dynamical critical exponent $z$,
whose value is controlled by $\kappa$.  When $|\kappa|\gg1$,
\begin{equation}\label{eq:zeff}
   z = 2 - \frac3{2|\kappa|} + O (|\kappa|^{-2}).
\end{equation}
Thus, the deep IR of UHCS gravity at sufficiently large $|\kappa|$ is a nontrivial example of a theory with a dynamical metric that shows scale-invariant behavior at the full quantum level.

\emph{Gauge symmetry.}---The algebra of the gauge symmetry of UHCS is the classical $w_\infty$ algebra, whose generators are functions of spatial coordinates and the commutators between the generators are their Poisson brackets $\{f, \, g\}=\epsilon^{ij}\d_i f\d_j g$. The gauge fields of UHCS are a scalar potential $A_0$ and an asymptotically flat Riemannian metric $g_{ij}$ on the spatial slices, which we take to have plane topology. We fix a reference Euclidean metric $\delta_{ij}$ and require $g_{ij}$ to be unimodular, meaning that its volume form coincides with that of $\delta_{ij}$, so 
$\det g=1$ in the inertial coordinates of $\delta_{ij}$.

Infinitesimal gauge transformations of the theory are parametrized by a scalar function $\alpha(t,\x)$, under which the fields transform as
\begin{subequations}\label{eq:gauge-transf}
\begin{align}
\delta_\alpha A_0 &= \d_t \alpha - \{ A_0, \, \alpha\}, \label{eq:gauge-transf-A0}\\
\delta_\alpha g_{ij} &=-\varepsilon^{kl}(\d_k g_{ij} \d_l \alpha + g_{kj}\d_i\d_l\alpha + g_{ik}\d_j \d_l\alpha). \label{eq:gauge-transf-gij}
\end{align}
\end{subequations}
Unimodularity of $g_{ij}$ is preserved under (\ref{eq:gauge-transf-gij}).  These transformation laws are motivated by the symmetries of the fractional quantum Hall effect~\cite{APD}.  Electrons in the lowest Landau level can be subjected to external probes $A_0$ and $g_{ij}$, coupled to the density and the stress tensor, respectively, in such a way that the partition function is invariant under (\ref{eq:gauge-transf}). Here, we promote the fields $A_0$ and $g_{ij}$ to dynamical gauge fields.

One can make a few additional comments on Eqs.\ (\ref{eq:gauge-transf}).  First, $A_0$ transforms like the temporal component of a gauge field.  The conventional way to construct a gauge theory is to extend $A_0$ to the full vector gauge field $A_\mu$ that transforms as $\delta_\alpha A_\mu=\d_\mu\alpha-\{A_\mu, \, \alpha\}$ and to build an action from the field strength $F_{\mu\nu}=\d_\mu A_\nu - \d_\nu A_\mu - \{A_\mu, \, A_\nu\}$.  Such a theory is the ``Poisson gauge theory'' \cite{Kupriyanov:2021aet}.  UHCS gravity is a different theory: in place of $A_i$, one has a tensor field $g_{ij}$ with a  nontrivial transformation law (\ref{eq:gauge-transf-gij}).  The connection to gravity becomes more transparent when one defines the vector $\xi^i=\epsilon^{ij}\d_j\alpha$ and rewrites Eq.\ (\ref{eq:gauge-transf-gij}) in terms of the Lie derivative $\mathcal{L}_{\xi}$ as
\begin{equation}
  \delta_\alpha g_{ij} = -(\mathcal L_\xi g)_{ij} \, ,
\end{equation}
which means that $g_{ij}$ transforms as a rank-2 tensor under the coordinate transformation $x^i\to x^i+\xi^i$. Note that $\d_i\xi^i=0$, so the coordinate transformation is an area-preserving diffeomorphism (APD) and $\alpha$ is the Hamiltonian generating this APD. The two transformation laws (\ref{eq:gauge-transf}) are consistent with each other and with the structure of the gauge group: their commutator is 
$[\delta_{\alpha_1}, \, \delta_{\alpha_2}] = \delta_{-\{\alpha_1,\, \alpha_2\}}$
when acting on either $A_0$ or $g_{ij}$.

\emph{Unimodular Ho\v rava gravity.}---To write down the action, we first work out the quantities (operators) $O$ that transform as
\begin{equation}
  \delta_\alpha O = -\{O, \, \alpha\} = -\xi^k\d_k O
\end{equation}
under the gauge transformation (\ref{eq:gauge-transf}).
One can either say that $O$ transforms in the adjoint representation of $w_\infty$ or like a scalar under APDs.  Such quantities can be used to construct the action, since $\int d^3x\, O$ is gauge invariant. The simplest operator contributing to the potential energy is $R^2$ (as $R\,d^2x$ is an exact differential). The leading-order potential energy $\int d^2x\, R^2$ is the Calabi energy of the 2D manifold. 
To construct the kinetic energy, one can check that the following object
\begin{equation}
  K_{ij} = \frac12 \left( \dot  g_{ij}-\nabla_i N_j - \nabla_j N_i\right),
  \qquad N^i = -\epsilon^{ij}\d_j A_0,
\end{equation}
transforms as a tensor under the gauge transformation: $\delta K_{ij}=-(\mathcal L_\xi K)_{ij}$.  Thus, the simplest version of the gravity theory has the action
\begin{equation}\label{eq:UH}
  S = \int\!d^3x \left( \frac4{C_1^2} K^{ij} K_{ij} - \frac1{C_2^2} R^2\right),
\end{equation}
where $C_1,\, C_2>0$.
We call this theory \emph{unimodular Ho\v rava gravity}.

Compared with projectable Ho\v rava gravity \cite{Horava:2009uw}, this theory has a divergenceless shift $N^i$ and a fixed lapse $N=1$.
Unimodular Ho\v rava gravity is the limit when one coupling constant of projectable Ho\v rava gravity (usually denoted as $\lambda$) is taken to infinity \cite{Radkovski:2023cew} with the Hilbert-space constraint of a zero total Hamiltonian removed.

\emph{Power counting.}---One can parametrize the unimodular metric by
\begin{equation}\label{eq:introh}
    g_{ij}=\delta_{ik} \left[\exp (h)\right]^k_{\:\:j},
\end{equation}
where $h^i_{\:\:j}$ is a symmetric traceless linear operator, i.e., $\delta^{ik}\delta_{jl} h^l_{\:\:k}=h^i_{\:\:j}$ and $h^i_{\:\:i}=0$. With $h_{ij}=\delta_{ik}h^k_{\:\:j}$, the action (\ref{eq:UH}) at the quadratic level reads
\begin{multline}\label{eq:Squad}
  S = \int\!d^3x\biggl[ \frac1{C_1^2}\bigl(\dot h_{ij}+\epsilon_{ik}\d_j\d_k A_0 +\epsilon_{jk}\d_i\d_k A_0\bigr)^2 \\ - \frac1{C_2^2}\left(\d_i \d_j h_{ij}\right)^2 \biggr],
\end{multline}
which implies a propagating graviton with a quadratic dispersion relation $\omega\propto |\vec k|^2$.  As noted in Ref.\ \cite{APD}, the quadratic theory is equivalent to a higher-rank gauge theory called ``traceless scalar charge theory'' \cite{Pretko:2016kxt,Pretko:2016lgv}, up to a change of variables.

Not only the quadratic action, but also the fully nonlinear classical theory is invariant under the anisotropic rescaling $t\to s^2 t$, $\x \to s \x$. The fields $A_0$ and $g_{ij}$ and the coupling constants $C_1$ and $C_2$ are dimensionless in the nonrelativistic power-counting scheme ($[x]=-1$, $[t]=-2$).  Thus, the theory (\ref{eq:UH}) is power-counting renormalizable.  It is straightforward to show that $C_1 C_2$ is the expansion parameter of perturbation theory.

\emph{One-loop running in unimodular Ho\v rava gravity.}---We now compute the one-loop running of the coupling constants using the background field method. Instead of Eq.\ (\ref{eq:introh}), in this calculation we will use the parametrization
\begin{equation}
  g(\Phi_1,\Phi_2) = \begin{pmatrix}
  e^{\Phi_1}\sqrt{1+\Phi_2^2} & \Phi_2\\ \Phi_2 & e^{-\Phi_1}\sqrt{1+\Phi_2^2}
  \end{pmatrix},
\end{equation}
for the unimodular metric in terms of two fields $\Phi_1$ and $\Phi_2$. This parametrization is convenient because the path integral measure $\mathcal D g_{ij}$ induced by any member of the DeWitt family of supermetrics \cite{DeWitt} is proportional to $\mathcal D\Phi_1\,\mathcal D \Phi_2$.  We now decompose the fields into backgrounds (marked by tildes) and fluctuations (written in lowercase):
\begin{subequations}
\begin{align}
  \Phi_n &= \widetilde\Phi_n +\phi_n, \quad n=1,2,\\
  A_0 &= \widetilde A_0+ a_0.
\end{align}
\end{subequations}
We also let $\tilde g=g(\widetilde\Phi_1,\widetilde\Phi_2)$ and $\gamma=g(\Phi_1,\Phi_2)-\tilde g$.

We calculate the one-loop contribution to the effective action on the background
\begin{equation}\label{eq:background}
\widetilde\Phi_1 = b(t,x,y), \qquad \widetilde\Phi_2 = \widetilde A_0 = 0.
\end{equation}
The gauge-fixing term we choose to add to the Lagrangian density is
\begin{multline}\label{eq:Lgf}
  \mathcal L_\text{gf}= -\frac1{2\xi} \bigg[ \frac2{C_1^2}\left(\d_0 a_0+\epsilon^{ij}\d_i a_0 \d_j \widetilde{A}_0\right) + \frac1{C_2^2}\epsilon^{ki} \widetilde\nabla_k \widetilde\nabla^j\gamma_{ij}\bigg]^2\\
  =-\frac1{2\xi} \left( \frac2{C_1^2}\d_0 a_0 + \frac1{C_2^2}\epsilon^{ki} \widetilde\nabla_k \widetilde\nabla^j\gamma_{ij}\right)^2,
\end{multline}
where $\widetilde\nabla$ is compatible with $\tilde g_{ij}$ through $\widetilde\nabla_{\!i}\tilde g_{jk}=0$ and $\xi$ is an arbitrary gauge parameter. This term fixes the gauge in the sense of breaking quantum gauge invariance, but preserves background gauge invariance. At $\xi=(2C_2^2)^{-1}$, the mixed propagators among $a_0$, $\phi_1$, and $\phi_2$ vanish, and the remaining momentum-space Feynman propagators are particularly simple:
\begin{subequations}
    \begin{align}
    \wick{\c{\widehat{a}}_0(p_{\mu}) \c{\widehat{a}}_0(q_{\mu})}&=-\frac{i C_1^4}{8 C_2^2 p_0^2-4 C_1^2 | \vec{p}|^4+i\epsilon}(2\pi)^3\delta^{(3)}(p+q)\,,\\
        \wick[offset=3ex]{\c{\widehat{\phi}}_1(p_{\mu}) \c{\widehat{\phi}}_1(q_{\mu})}&=\wick[offset=3ex]{\c{\widehat{\phi}}_2(p_{\mu}) \c{\widehat{\phi}}_2(q_{\mu})}\nonumber\\
        &=\frac{i C_1^2 C_2^2}{4 C_2^2 p_0^2-2 C_1^2 | \vec{p}|^4+i\epsilon}(2\pi)^3\delta^{(3)}(p+q)\,,
    \end{align}
\end{subequations}
where $\widehat{f}(p_{\mu})=\int dt\, d^2x\, f(x^{\mu}) e^{i (p_0 t- p_i x^i)}$
for any field $f$. We introduce the Faddeev--Popov ghosts $c$ and $\overline{c}$, along with the ghost term
\begin{multline}\label{eq:Lgh}
    \mathcal{L}_{\rm gh} = -\overline{c}\biggl[\frac{2}{C_1^2}\left(\partial_0^2 c-\epsilon^{ij}\left(\partial_0 \partial_i a_0 \partial_j c+\partial_i a_0 \partial_0\partial_j c\right)\right) \\ +\frac{1}{C_2^2}\widetilde{\nabla}_k\widetilde{\nabla}_l M^{kl}\left(\tilde{g}_{ij},\gamma_{ij},c\right)\biggr],
\end{multline}
where
\begin{multline}
    M^{kl}\left(\tilde{g}_{ij},\gamma_{ij},c\right):=\tilde{g}^{\,lj}\\\times\left[\left(\widetilde{\nabla}_m\gamma_{ij}\right)+g_{mj}\widetilde{\nabla}_{i}+g_{mi}\widetilde{\nabla}_{j}\right]\left(\tilde{g}^{im}\widetilde{\nabla}^k c-\tilde{g}^{km}\widetilde{\nabla}^i c\right).
\end{multline}

By Eq.\ (\ref{eq:Squad}) for 
$h=\text{diag}(b,-b)$, the quadratic part of the action evaluated at our background (\ref{eq:background}) is
\begin{equation}\label{eq:backgroundaction}
    S=\int\! dt\, d^2x\left[\frac{2}{C_1^2}(\partial_0 b)^2-\frac{1}{C_2^2}(\partial_1^2 b-\partial_2^2 b)^2\right].
\end{equation}
Accordingly, the one-loop quantum effective action in momentum space has the form
\begin{multline}\label{eq:momGamma}
    \Gamma=\int \frac{dk_0\,d^2 k}{(2\pi)^3}\,\widehat{b}(-k_{\mu})\widehat{b}(k_{\mu})\left[\frac{2k_0^2}{C_1^2}-\frac{\left(k_1^2-k_2^2\right)^2}{C_2^2}\right.\\
    \left.-i\left(\raisebox{-0.15 in}{\includegraphics[height=0.5 in]{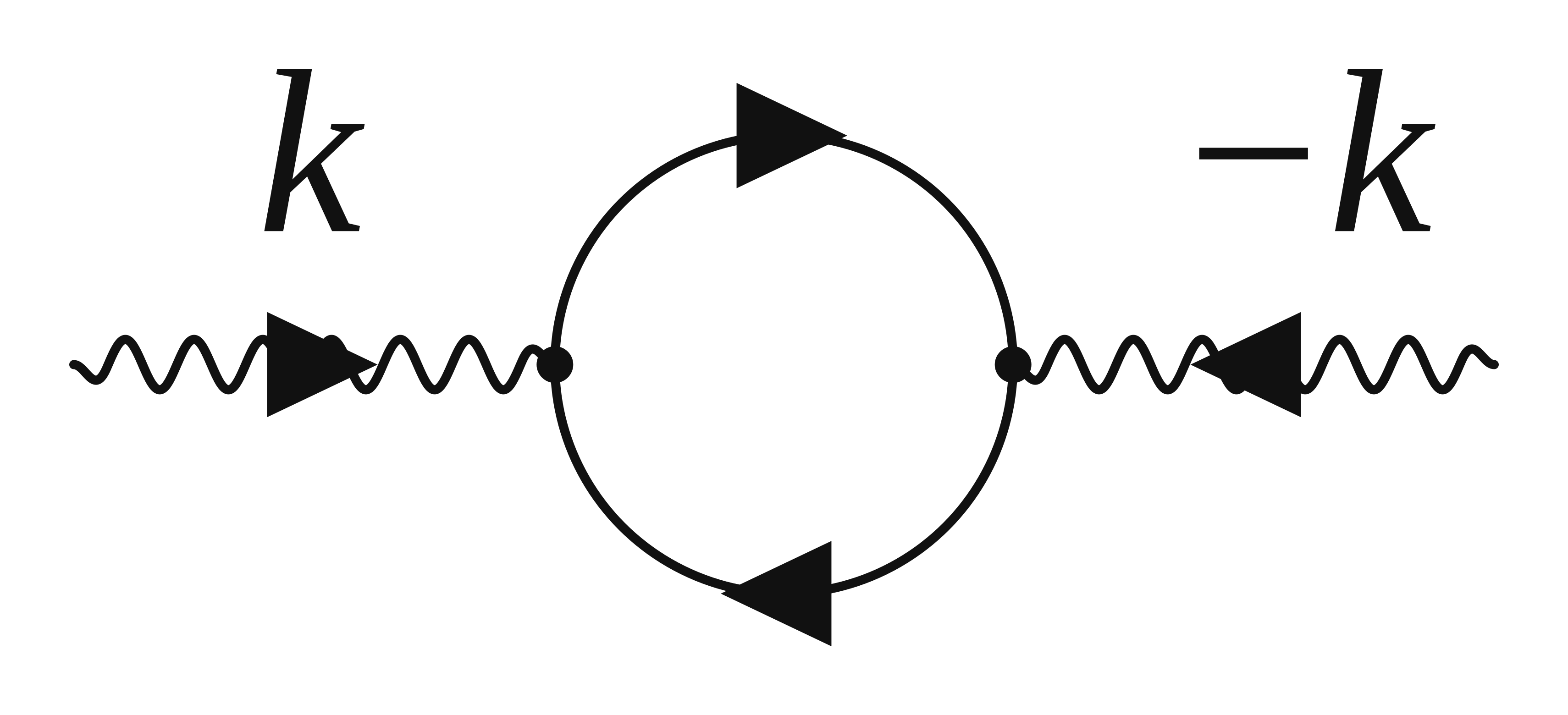}}+\raisebox{-0.2 in}{\includegraphics[height=0.5 in]{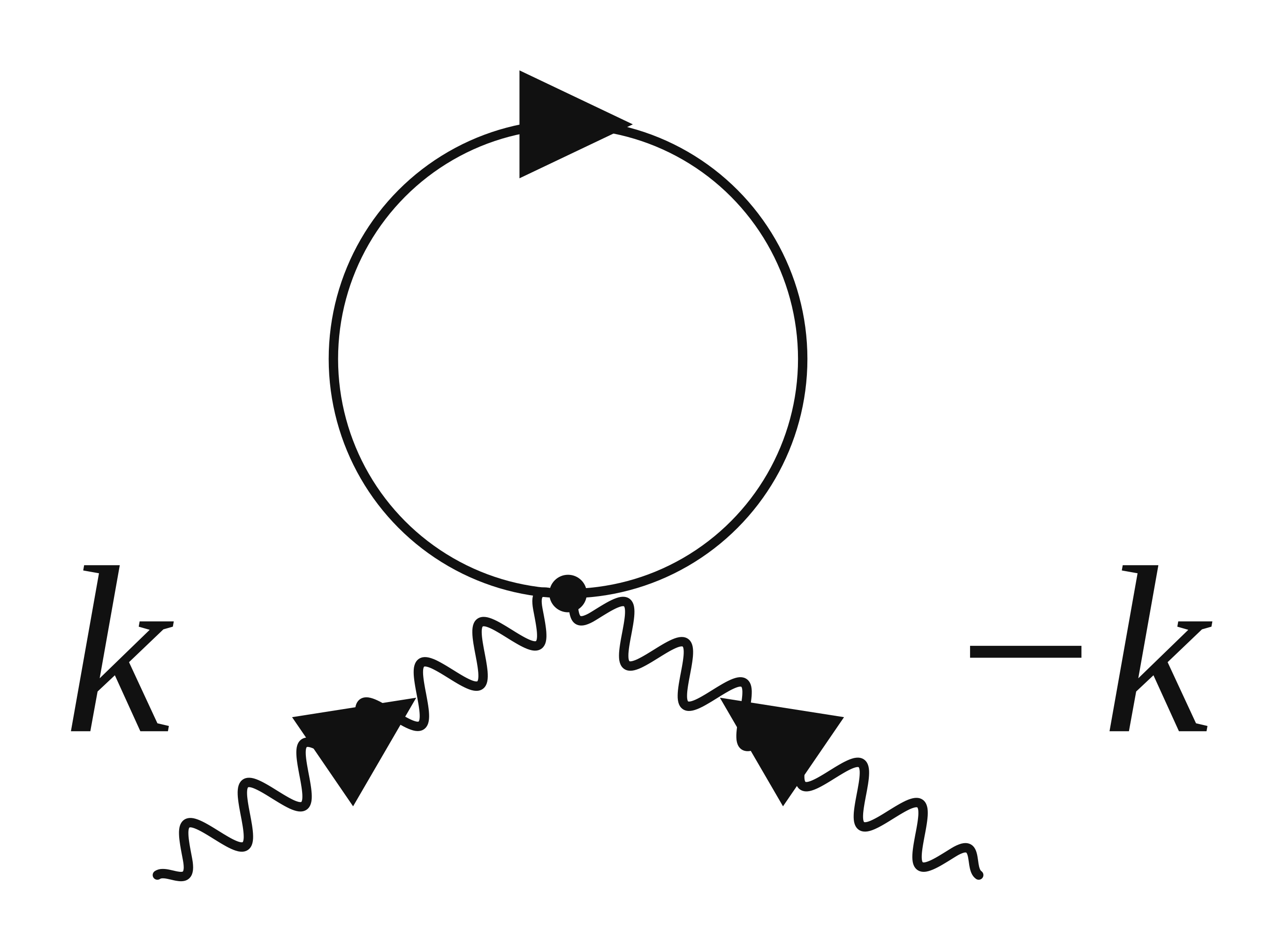}}\right)\right]
\end{multline}
to quadratic order in $b$, where the Feynman diagrams account for contributions from the original Lagrangian density, the gauge-fixing term (\ref{eq:Lgf}), and the ghost term (\ref{eq:Lgh}). In these diagrams, each wavy line represents a Fourier mode of the background field $b$ with external momentum $k_{\mu}$, and each plain line is a propagator of the fluctuations $a_0$, $\phi_1$, $\phi_2$, $c$, and $\overline{c}$, with the boson loops carrying a factor of $1/2$ and the ghost loops carrying a minus sign. By converting the propagators and vertices to momentum space, we can write the diagrams as integrals over a single loop momentum $p_{\mu}$.

Using Mathematica \cite{Mathematica}, we compute the net contribution of these diagrams for arbitrary values of $\xi$ \footnote{The computations are carried out in the Mathematica notebook \texttt{RG\_UHG\_without\_CS.nb}, which is included as an ancillary file with the arXiv version of this paper.}. We use the package GRQUICK~\cite{GRQUICK} to find the terms in the Lagrangian density needed for the Feynman rules. We then implement those rules to set up the integral over $p_{\mu}$.
To integrate over the loop momentum $p_{\mu}$, we rescale $k_0\to s^2 k_0$ and $\vec{k}\to s\vec{k}$, Taylor expand to fourth order in $s$ (the order of $k_0^2$ and $|\vec{k}|^4$), set $s=1$, remove the part of the integrand odd in $p_0$, average over the direction of the spatial loop momentum $\vec{p}$, and integrate over $p_0$ after a Wick rotation. The remaining integral over $|\vec{p}|$ is evaluated in dimensional regularization, which discards the pure power-law UV divergences and leaves a logarithm. The logarithm's UV divergence is removed by modified minimal subtraction at the scale $M$, while its IR end is cut off by the external momentum scale $\mu$, since the Taylor expansion caused the integrand's denominators to lose their dependence on the external momentum, which regulates the IR of the unexpanded integral.
\par As expected from the background gauge invariance of $\Gamma$, the result is a linear combination of $k_0^2$ and $\left(k_1^2-k_2^2\right)^2$:
\begin{multline}
    -i\left(\raisebox{-0.15 in}{\includegraphics[height=0.5 in]{VPDbubble.png}}+\raisebox{-0.2 in}{\includegraphics[height=0.5 in]{VPDseagull.png}}\right)\\
    =\left(-\frac{7 C_2 k_0^2}{32 \sqrt{2} \pi  C_1}+\frac{C_1 \left(k_1^2-k_2^2\right)^2}{32 \sqrt{2} \pi  C_2}\right)\ln\left(\frac{M}{\mu}\right).
\end{multline}
These one-loop corrections correspond to the beta functions
\begin{subequations}\label{eq:RGeq}
\begin{align}
    \beta_{C_1}=\mu \frac{{\mathrm d} C_1}{{\mathrm d}\mu}&=-\frac{7 C_1^2 C_2}{128\sqrt{2}\pi} \,, \label{eq:RGeqC1}\\
    \beta_{C_2}= \mu \frac{{\mathrm d} C_2}{{\mathrm d}\mu}&=-\frac{C_1 C_2^2}{64\sqrt{2}\pi} \,. \label{eq:RGeqC2}
\end{align}
\end{subequations}
In particular,
\begin{equation}
    \mu \frac{{\mathrm d}}{{\mathrm d}\mu}\left(C_1 C_2\right) = -\frac{9(C_1 C_2)^2}{128\sqrt{2}\pi}\,,\label{eq:RGeqC1C2}
\end{equation}
implying that we have an asymptotically free theory.  Taking the ratio of the two equations (\ref{eq:RGeqC1}) and (\ref{eq:RGeqC2}), we find that $C_2^7/C_1^2$ remains constant along the RG flow. 
 
Equation (\ref{eq:RGeqC1C2}) matches the $\lambda\to\infty$ limit of the running of the coupling $C_1 C_2$ in (2+1)D projectable Ho\v rava gravity \cite{Sibiryakov}, which is consistent with the aforementioned fact that unimodular Ho\v rava gravity is the $\lambda\to\infty$ limit with the global Hamiltonian constraint on the Hilbert space removed, since that constraint does not affect the local divergences that make the couplings run. 
While
$C_1C_2$ is the only physical coupling in projectable Ho\v rava gravity \cite{Sibiryakov}, the couplings $C_1$ and $C_2$ are separately physical in unimodular Ho\v rava gravity because there is no lapse and $\det g=1$.

\emph{CS terms.}---The two terms in the action (\ref{eq:UH}) are the leading quantities transforming like scalars under APDs---any other scalar will have a higher scaling dimension and be irrelevant. On the other hand, the term
\begin{equation}\label{eq:CS1}
    \mathcal L_\text{CS1}=\rho A_0,
\end{equation}
where $\rho$ is a coupling, is a Chern--Simons term in the sense that it is independent of the unimodular metric and fails to be a scalar under APDs only through the total derivative $\d_t \alpha$ in the transformation (\ref{eq:gauge-transf-A0}). Such a CS term integrates to an APD-invariant contribution to the action modulo a boundary term at $t\rightarrow\pm\infty$. Being linear in the fields, $\mathcal L_\text{CS1}$ acts as a source term, not contributing to the Feynman rules for the 1PI diagrams in the quantum effective action, so $\rho$ neither runs nor affects the running of the other couplings. It only modifies the Gauss law phase-space constraint obtained by varying the action with respect to $A_0$, so it changes the physical Hilbert space.
\par More interesting CS terms arise if we introduce a spatial vielbein $e^a_i$ such that $g_{ij}=\delta_{ab} e^a_i e^b_j$ and $\epsilon_{ij}=\epsilon_{ab} e^a_i e^b_j$, because we gain a new internal SO(2) gauge symmetry consisting of local vielbein rotations $e^a_i\rightarrow R^a_{\:\:b}(\theta) e^b_i$ that leave the metric unchanged. Letting $e^{ai}=g^{ij}e^a_j$, we can add two other CS terms that are scalars under APDs and vary only by exact differentials under the internal SO(2):
\begin{align}
  \mathcal L_\text{CS2} &= \eta\left( \omega_0 -N^i\omega_i \right)\label{eq:CS2}\\
  \mathcal L_\text{CS3} &= \frac{\kappa}{4\pi} \epsilon^{\mu\nu\lambda}\omega_\mu \d_\nu \omega_\lambda,\qquad \epsilon^{0ij}=\epsilon^{ij},\label{eq:CS3}
\end{align}
where $\eta$ and $\kappa$ are coefficients of the CS terms and
\begin{subequations}
\begin{align}
    \omega_0 &=\frac12 \left(\epsilon^{ab} e^{aj}\d_0 e^b_j - \epsilon^{ij}\partial_i N_j\right)\nonumber\\
    &= \frac12 \left(\epsilon^{ab} e^{aj}\d_0 e^b_j - \nabla^2 A_0\right), \\
    \omega_i &= \frac12 \epsilon^{ab} e^{aj}\nabla_i e^b_j
\end{align}
\end{subequations}
are the time and space components of an SO(2) spin connection of the spatial vielbein.

Power counting shows that $\mathcal L_\text{CS2}$ is a dimension-2 operator and is relevant. Adding this term to the action gaps out the graviton. We assume that $\rho=\eta=0$ in the terms (\ref{eq:CS1}) and (\ref{eq:CS2}). This assumption is stable under RG flow, since $\rho$ does not run, as explained above, while $\eta$ does not run because it is a CS coefficient quantized in units of the inverse area of closed, oriented spatial slices, of which our slices are infinite-area limits. We thus add to the theory only the third CS term. The full action is now
\begin{equation}\label{eq:UHCS}
  S = \int\!d^3x \left( \frac4{C_1^2} K^{ij} K_{ij} - \frac1{C_2^2} R^2 + \frac{\kappa}{4\pi} \epsilon^{\mu\nu\lambda}\omega_\mu \d_\nu \omega_\lambda \right).
\end{equation}
Every coupling constant is dimensionless, so the theory remains power-counting renormalizable.  The theory (\ref{eq:UHCS}) is what we call the unimodular Ho\v rava--Chern--Simons (UHCS) gravity.  At the quadratic level, Eq.\ (\ref{eq:UHCS}) describes a graviton with the dispersion relation
\begin{equation}\label{eq:dispersion}
  \omega^2 = \frac{C_1^2}{4 C_2^2} 
  \left(2+\widetilde{\kappa}^2\right) |\vec{k}|^4, \qquad \widetilde{\kappa}=\frac{\kappa C_1 C_2}{16\pi}.
\end{equation}

As a Chern--Simons level, $\kappa$ does not run, but it influences the beta functions of the other coupling constants.  Using Mathematica \footnote{The computations are carried out in the Mathematica notebook \texttt{RG\_UHCS\_gravity.nb}, which is included as an ancillary file with the arXiv version of this paper.}, we repeat the calculation of the beta functions with $\kappa\neq0$ for the background (\ref{eq:background}), but in a different background field gauge where we replace
\begin{equation}
    \frac1{C_2^2}\rightarrow\frac1{C_2^2}\left(1+\frac{\widetilde{\kappa}^2}2\right)
\end{equation}
in both the gauge-fixing term (\ref{eq:Lgf}) and the ghost term (\ref{eq:Lgh}) and work only at $\xi=(2C_2^2)^{-1}(1+{\widetilde{\kappa}^2}/2)$. The internal SO(2) symmetry is gauge-fixed by $e^2_1=0$, so the vielbein becomes
\begin{equation}
    e(\Phi_1,\Phi_2) = \begin{pmatrix}
  e^{\Phi_1/2}\left(1+\Phi_2^2\right)^{1/4} & \frac{\Phi_2 e^{-\Phi_1/2}}{\left(1+\Phi_2^2\right)^{1/4}}\\ 0 & \frac{e^{-\Phi_1/2}}{\left(1+\Phi_2^2\right)^{1/4}}
  \end{pmatrix}.
\end{equation}

The resulting RG equations are
\begin{subequations}\label{eq:RG-with-CS}
\begin{align}
  \mu \frac{{\mathrm d}C_1}{\mathrm d\mu} &= -C_1^2 C_2 \frac{7+3\widetilde\kappa^2}{64\pi (2+\widetilde\kappa^2)^{3/2}}, \label{eq:RG-with-CS1}\\
  \mu \frac{{\mathrm d}C_2}{\mathrm d\mu} &= -C_1 C_2^2 \frac1{32\pi (2+\widetilde\kappa^2)^{3/2}},
  \label{eq:RG-with-CS2}
\end{align}
\end{subequations}
so
\begin{subequations}\label{eq:useful-CS-RG}
    \begin{align}
    \mu \frac{{\mathrm d}\left|\widetilde{\kappa}\right|}{\mathrm d\mu}&=\sgn(\kappa)\mu \frac{{\mathrm d}\widetilde{\kappa}}{\mathrm d\mu}=-\widetilde{\kappa}^2\frac{9+3\widetilde\kappa^2}{4|\kappa|(2+\widetilde\kappa^2)^{3/2}},\label{eq:prodRG}\\
    \frac{{\mathrm d}\ln \left|\widetilde{\kappa}\right|}{{\mathrm d}\ln C_2}&=1+\frac{{\mathrm d}\ln C_1}{{\mathrm d}\ln C_2}=\frac{9+3\widetilde{\kappa}^2}2. \label{eq:ratioRG}
\end{align}
\end{subequations}
The right-hand side of Eq.\ (\ref{eq:prodRG}) is negative, so $|\widetilde\kappa|$ grows without bound as $\mu$ decreases in the direction of RG flow. This growth causes the right-hand side to asymptote to ${-3\left|\widetilde\kappa\right|/(4|\kappa|)}$, so ${\left|\widetilde{\kappa}(\mu)\right|\propto \mu^{-\frac{3}{4|\kappa|}}}$ as $\mu\rightarrow 0$. Flowing back to the UV, $|\widetilde\kappa|$ instead decreases to zero, turning Eqs.\ (\ref{eq:RG-with-CS}) into Eqs.\ (\ref{eq:RGeq}), thereby recovering the asymptotic freedom of the $\kappa=0$ case.

By separation of variables, Eq.\ (\ref{eq:ratioRG}) implies that the RG curves have constant ${C_2^9(1+3/\widetilde\kappa^2)}\propto {C_2^7(1+\widetilde\kappa^2/3)/C_1^2}$. The latter constant reduces to ${C_2^7/C_1^2}$ for $\kappa=0$, whereas the former implies that $C_2$ increases toward a finite value as RG makes ${\left|\widetilde\kappa\right|\rightarrow\infty}$. Thus, the blowup $\left|\widetilde{\kappa}(\mu)\right|\propto \mu^{-\frac{3}{4|\kappa|}}$ must be solely due to $C_1\rightarrow\infty$, so $C_1(\mu)\propto \mu^{-\frac{3}{4|\kappa|}}$ as $\mu\rightarrow 0$. This power law turns the dispersion relation (\ref{eq:dispersion}) into
\begin{equation}
    \omega \propto \bigl[C_1(|\vec{k}|)\bigr]^2|\vec{k}|^2 \propto |\vec{k}|^{2-\frac{3}{2|\kappa|}},
\end{equation}
implying the dynamical critical exponent (\ref{eq:zeff}) \footnote{Had we instead set $\mu=\sqrt{\omega}$ in $C_1(\mu)$, we would have derived 
$z=\frac2{1+3/(4|\kappa|)}={2-\frac{3}{2|\kappa|}+O (|\kappa|^{-2})}$, which differs from the $\mu=|\vec{k}|$ result only by $O(|\kappa|^{-2})$, beyond the accuracy of a one-loop calculation.}.

The fact that the IR behavior corresponds to a Lifshitz fixed point can be seen more clearly by noticing that, if one changes the RG procedure to correspond to RG steps $\mathbf x\to s \mathbf x$, $t\to s^{z}t$ (instead of $s^2t$), then, for $z\approx2$, the right-hand sides of Eqs.\ (\ref{eq:RG-with-CS1}) and (\ref{eq:RG-with-CS2}) acquire the extra terms $(1-z/2)C_1$ and $-(1-z/2)C_2$, respectively. Choosing $z=2-\frac3{2|\kappa|}$ then yields the fixed point $C_1=\text{const}$, $1/C_2^{2}=0$ \footnote{Different values of $C_1$ at the fixed point are equivalent through time rescaling $t\to\lambda t$.  At the one-loop level, the fixed point corresponds to a vanishing coefficient $1/C_2^2$ of the Calabi energy term, or equivalently, to $\protect\widetilde{\kappa}^2=\infty$.  At the two- and higher-loop levels, it is possible that the fixed point moves to a nonzero $1/C_2^2$, i.e., a finite $\protect\widetilde{\kappa}^2$.  As long as $\protect\widetilde{\kappa}^2 \gg1$ at the fixed point (which should be the case for $|\kappa|\gg1$), our prediction for the dynamical critical exponent, obtained by demanding the vanishing of the beta function for $C_1$, still stands.}.

\emph{Conclusions.}--- For $|\kappa|\gg 1$, UHCS gravity is a theory where the one-loop RG flow is from an asymptotically free UV fixed point with $z=2$, where the quadratic part of the action without the CS term dominates, to an IR fixed point with a dynamical critical exponent (\ref{eq:zeff}). 

One interesting future direction is to determine whether or not there exists a phase transition at a critical $|\kappa|=\kappa_*$, above which the IR Lifshitz fixed point is reached and below which the theory flows to strong coupling in the IR as for $\kappa=0$, possibly into a confining phase.
For $e^{iS}$ to be invariant under large internal SO(2) gauge transformations on closed, oriented spatial slices, $\kappa$ must be an integer.
However, by coupling UHCS gravity to dynamical U(1) Chern--Simons fields through $a\wedge d\omega$-type couplings, as in the $K$-matrix theories of the fractional quantum Hall effect, one may be able to achieve effective rational values of $\kappa$.

\acknowledgments

We thank Petr Ho\v rava and Alexander Vikman for valuable comments.  This work is supported, in part, by the U.S.\ DOE Grant No.\ DE-FG02-13ER41958, by the Simons Collaboration on Ultra-Quantum Matter, which is a grant from the Simons Foundation (Grant No.\ 651442, D.T.S.).  The authors used Claude (Fable 5, Fable 5.1, and Opus 5.5) and ChatGPT-6 Astra for checking the results, most notably the independence of the one-loop beta functions from the gauge-fixing parameter.

\bibliography{VPDbib}

\begin{thebibliography}{17}%
\makeatletter
\providecommand \@ifxundefined [1]{%
 \@ifx{#1\undefined}
}%
\providecommand \@ifnum [1]{%
 \ifnum #1\expandafter \@firstoftwo
 \else \expandafter \@secondoftwo
 \fi
}%
\providecommand \@ifx [1]{%
 \ifx #1\expandafter \@firstoftwo
 \else \expandafter \@secondoftwo
 \fi
}%
\providecommand \natexlab [1]{#1}%
\providecommand \enquote  [1]{``#1''}%
\providecommand \bibnamefont  [1]{#1}%
\providecommand \bibfnamefont [1]{#1}%
\providecommand \citenamefont [1]{#1}%
\providecommand \href@noop [0]{\@secondoftwo}%
\providecommand \href [0]{\begingroup \@sanitize@url \@href}%
\providecommand \@href[1]{\@@startlink{#1}\@@href}%
\providecommand \@@href[1]{\endgroup#1\@@endlink}%
\providecommand \@sanitize@url [0]{\catcode `\\12\catcode `\$12\catcode
  `\&12\catcode `\#12\catcode `\^12\catcode `\_12\catcode `\%12\relax}%
\providecommand \@@startlink[1]{}%
\providecommand \@@endlink[0]{}%
\providecommand \url  [0]{\begingroup\@sanitize@url \@url }%
\providecommand \@url [1]{\endgroup\@href {#1}{\urlprefix }}%
\providecommand \urlprefix  [0]{URL }%
\providecommand \Eprint [0]{\href }%
\providecommand \doibase [0]{https://doi.org/}%
\providecommand \selectlanguage [0]{\@gobble}%
\providecommand \bibinfo  [0]{\@secondoftwo}%
\providecommand \bibfield  [0]{\@secondoftwo}%
\providecommand \translation [1]{[#1]}%
\providecommand \BibitemOpen [0]{}%
\providecommand \bibitemStop [0]{}%
\providecommand \bibitemNoStop [0]{.\EOS\space}%
\providecommand \EOS [0]{\spacefactor3000\relax}%
\providecommand \BibitemShut  [1]{\csname bibitem#1\endcsname}%
\let\auto@bib@innerbib\@empty
\bibitem [{\citenamefont {Goroff}\ and\ \citenamefont
  {Sagnotti}(1986)}]{Goroff:1985th}%
  \BibitemOpen
  \bibfield  {author} {\bibinfo {author} {\bibfnamefont {M.~H.}\ \bibnamefont
  {Goroff}}\ and\ \bibinfo {author} {\bibfnamefont {A.}~\bibnamefont
  {Sagnotti}},\ }\bibfield  {title} {\bibinfo {title} {{The Ultraviolet
  Behavior of Einstein Gravity}},\ }\href
  {https://doi.org/10.1016/0550-3213(86)90193-8} {\bibfield  {journal}
  {\bibinfo  {journal} {Nucl. Phys. B}\ }\textbf {\bibinfo {volume} {266}},\
  \bibinfo {pages} {709} (\bibinfo {year} {1986})}\BibitemShut {NoStop}%
\bibitem [{\citenamefont {Ho{\v{r}}ava}(2009)}]{Horava:2009uw}%
  \BibitemOpen
  \bibfield  {author} {\bibinfo {author} {\bibfnamefont {P.}~\bibnamefont
  {Ho{\v{r}}ava}},\ }\bibfield  {title} {\bibinfo {title} {{Quantum Gravity at
  a Lifshitz Point}},\ }\href {https://doi.org/10.1103/PhysRevD.79.084008}
  {\bibfield  {journal} {\bibinfo  {journal} {Phys. Rev. D}\ }\textbf {\bibinfo
  {volume} {79}},\ \bibinfo {pages} {084008} (\bibinfo {year} {2009})},\
  \Eprint {https://arxiv.org/abs/0901.3775} {arXiv:0901.3775 [hep-th]}
  \BibitemShut {NoStop}%
\bibitem [{\citenamefont {Barvinsky}\ \emph {et~al.}(2017)\citenamefont
  {Barvinsky}, \citenamefont {Blas}, \citenamefont {Herrero-Valea},
  \citenamefont {Sibiryakov},\ and\ \citenamefont {Steinwachs}}]{Sibiryakov}%
  \BibitemOpen
  \bibfield  {author} {\bibinfo {author} {\bibfnamefont {A.~O.}\ \bibnamefont
  {Barvinsky}}, \bibinfo {author} {\bibfnamefont {D.}~\bibnamefont {Blas}},
  \bibinfo {author} {\bibfnamefont {M.}~\bibnamefont {Herrero-Valea}}, \bibinfo
  {author} {\bibfnamefont {S.~M.}\ \bibnamefont {Sibiryakov}},\ and\ \bibinfo
  {author} {\bibfnamefont {C.~F.}\ \bibnamefont {Steinwachs}},\ }\bibfield
  {title} {\bibinfo {title} {{Ho{\v{r}}ava Gravity is Asymptotically Free in 2
  + 1 Dimensions}},\ }\href {https://doi.org/10.1103/PhysRevLett.119.211301}
  {\bibfield  {journal} {\bibinfo  {journal} {Phys. Rev. Lett.}\ }\textbf
  {\bibinfo {volume} {119}},\ \bibinfo {pages} {211301} (\bibinfo {year}
  {2017})},\ \Eprint {https://arxiv.org/abs/1706.06809} {arXiv:1706.06809
  [hep-th]} \BibitemShut {NoStop}%
\bibitem [{\citenamefont {Barvinsky}\ \emph {et~al.}(2022)\citenamefont
  {Barvinsky}, \citenamefont {Kurov},\ and\ \citenamefont
  {Sibiryakov}}]{Barvinsky:2021ubv}%
  \BibitemOpen
  \bibfield  {author} {\bibinfo {author} {\bibfnamefont {A.~O.}\ \bibnamefont
  {Barvinsky}}, \bibinfo {author} {\bibfnamefont {A.~V.}\ \bibnamefont
  {Kurov}},\ and\ \bibinfo {author} {\bibfnamefont {S.~M.}\ \bibnamefont
  {Sibiryakov}},\ }\bibfield  {title} {\bibinfo {title} {{Beta functions of
  (3+1)-dimensional projectable Ho{\v{r}}ava gravity}},\ }\href
  {https://doi.org/10.1103/PhysRevD.105.044009} {\bibfield  {journal} {\bibinfo
   {journal} {Phys. Rev. D}\ }\textbf {\bibinfo {volume} {105}},\ \bibinfo
  {pages} {044009} (\bibinfo {year} {2022})},\ \Eprint
  {https://arxiv.org/abs/2110.14688} {arXiv:2110.14688 [hep-th]} \BibitemShut
  {NoStop}%
\bibitem [{\citenamefont {Barvinsky}\ \emph {et~al.}(2025)\citenamefont
  {Barvinsky}, \citenamefont {Kurov},\ and\ \citenamefont
  {Sibiryakov}}]{Barvinsky:2024svc}%
  \BibitemOpen
  \bibfield  {author} {\bibinfo {author} {\bibfnamefont {A.~O.}\ \bibnamefont
  {Barvinsky}}, \bibinfo {author} {\bibfnamefont {A.~V.}\ \bibnamefont
  {Kurov}},\ and\ \bibinfo {author} {\bibfnamefont {S.~M.}\ \bibnamefont
  {Sibiryakov}},\ }\bibfield  {title} {\bibinfo {title} {{Renormalization group
  flow of projectable Ho{\v{r}}ava gravity in (3+1) dimensions}},\ }\href
  {https://doi.org/10.1103/PhysRevD.111.024030} {\bibfield  {journal} {\bibinfo
   {journal} {Phys. Rev. D}\ }\textbf {\bibinfo {volume} {111}},\ \bibinfo
  {pages} {024030} (\bibinfo {year} {2025})},\ \Eprint
  {https://arxiv.org/abs/2411.13574} {arXiv:2411.13574 [gr-qc]} \BibitemShut
  {NoStop}%
\bibitem [{\citenamefont {Du}\ \emph {et~al.}(2022)\citenamefont {Du},
  \citenamefont {Mehta}, \citenamefont {Nguyen},\ and\ \citenamefont
  {Son}}]{APD}%
  \BibitemOpen
  \bibfield  {author} {\bibinfo {author} {\bibfnamefont {Y.-H.}\ \bibnamefont
  {Du}}, \bibinfo {author} {\bibfnamefont {U.}~\bibnamefont {Mehta}}, \bibinfo
  {author} {\bibfnamefont {D.~X.}\ \bibnamefont {Nguyen}},\ and\ \bibinfo
  {author} {\bibfnamefont {D.~T.}\ \bibnamefont {Son}},\ }\bibfield  {title}
  {\bibinfo {title} {{Volume-preserving diffeomorphism as nonabelian
  higher-rank gauge symmetry}},\ }\href
  {https://doi.org/10.21468/SciPostPhys.12.2.050} {\bibfield  {journal}
  {\bibinfo  {journal} {SciPost Phys.}\ }\textbf {\bibinfo {volume} {12}},\
  \bibinfo {pages} {050} (\bibinfo {year} {2022})},\ \Eprint
  {https://arxiv.org/abs/2103.09826} {arXiv:2103.09826 [cond-mat.str-el]}
  \BibitemShut {NoStop}%
\bibitem [{\citenamefont {Kupriyanov}(2021)}]{Kupriyanov:2021aet}%
  \BibitemOpen
  \bibfield  {author} {\bibinfo {author} {\bibfnamefont {V.~G.}\ \bibnamefont
  {Kupriyanov}},\ }\bibfield  {title} {\bibinfo {title} {{Poisson gauge
  theory}},\ }\href {https://doi.org/10.1007/JHEP09(2021)016} {\bibfield
  {journal} {\bibinfo  {journal} {JHEP}\ }\textbf {\bibinfo {volume} {09}},\
  \bibinfo {pages} {016}},\ \Eprint {https://arxiv.org/abs/2105.14965}
  {arXiv:2105.14965 [hep-th]} \BibitemShut {NoStop}%
\bibitem [{\citenamefont {Radkovski}\ and\ \citenamefont
  {Sibiryakov}(2023)}]{Radkovski:2023cew}%
  \BibitemOpen
  \bibfield  {author} {\bibinfo {author} {\bibfnamefont {J.~I.}\ \bibnamefont
  {Radkovski}}\ and\ \bibinfo {author} {\bibfnamefont {S.~M.}\ \bibnamefont
  {Sibiryakov}},\ }\bibfield  {title} {\bibinfo {title} {{Scattering amplitudes
  in high-energy limit of projectable Ho{\v{r}}ava gravity}},\ }\href
  {https://doi.org/10.1103/PhysRevD.108.046017} {\bibfield  {journal} {\bibinfo
   {journal} {Phys. Rev. D}\ }\textbf {\bibinfo {volume} {108}},\ \bibinfo
  {pages} {046017} (\bibinfo {year} {2023})},\ \Eprint
  {https://arxiv.org/abs/2306.00102} {arXiv:2306.00102 [hep-th]} \BibitemShut
  {NoStop}%
\bibitem [{\citenamefont {Pretko}(2017{\natexlab{a}})}]{Pretko:2016kxt}%
  \BibitemOpen
  \bibfield  {author} {\bibinfo {author} {\bibfnamefont {M.}~\bibnamefont
  {Pretko}},\ }\bibfield  {title} {\bibinfo {title} {{Subdimensional Particle
  Structure of Higher Rank U(1) Spin Liquids}},\ }\href
  {https://doi.org/10.1103/PhysRevB.95.115139} {\bibfield  {journal} {\bibinfo
  {journal} {Phys. Rev. B}\ }\textbf {\bibinfo {volume} {95}},\ \bibinfo
  {pages} {115139} (\bibinfo {year} {2017}{\natexlab{a}})},\ \Eprint
  {https://arxiv.org/abs/1604.05329} {arXiv:1604.05329 [cond-mat.str-el]}
  \BibitemShut {NoStop}%
\bibitem [{\citenamefont {Pretko}(2017{\natexlab{b}})}]{Pretko:2016lgv}%
  \BibitemOpen
  \bibfield  {author} {\bibinfo {author} {\bibfnamefont {M.}~\bibnamefont
  {Pretko}},\ }\bibfield  {title} {\bibinfo {title} {{Generalized
  Electromagnetism of Subdimensional Particles: A Spin Liquid Story}},\ }\href
  {https://doi.org/10.1103/PhysRevB.96.035119} {\bibfield  {journal} {\bibinfo
  {journal} {Phys. Rev. B}\ }\textbf {\bibinfo {volume} {96}},\ \bibinfo
  {pages} {035119} (\bibinfo {year} {2017}{\natexlab{b}})},\ \Eprint
  {https://arxiv.org/abs/1606.08857} {arXiv:1606.08857 [cond-mat.str-el]}
  \BibitemShut {NoStop}%
\bibitem [{\citenamefont {DeWitt}(1967)}]{DeWitt}%
  \BibitemOpen
  \bibfield  {author} {\bibinfo {author} {\bibfnamefont {B.~S.}\ \bibnamefont
  {DeWitt}},\ }\bibfield  {title} {\bibinfo {title} {{Quantum Theory of
  Gravity. I. The Canonical Theory}},\ }\href
  {https://doi.org/10.1103/PhysRev.160.1113} {\bibfield  {journal} {\bibinfo
  {journal} {Phys. Rev.}\ }\textbf {\bibinfo {volume} {160}},\ \bibinfo {pages}
  {1113} (\bibinfo {year} {1967})}\BibitemShut {NoStop}%
\bibitem [{\citenamefont {{Wolfram Research, Inc.}}(2026)}]{Mathematica}%
  \BibitemOpen
  \bibfield  {author} {\bibinfo {author} {\bibnamefont {{Wolfram Research,
  Inc.}}},\ }\href@noop {} {\emph {\bibinfo {title} {{Mathematica, Version
  15.0.1}}}},\ \bibinfo {address} {Champaign, IL} (\bibinfo {year}
  {2026})\BibitemShut {NoStop}%
\bibitem [{Note1()}]{Note1}%
  \BibitemOpen
  \bibinfo {note} {The computations are carried out in the Mathematica notebook
  \protect \texttt {RG\protect \_UHG\protect \_without\protect \_CS.nb}, which
  is included as an ancillary file with the arXiv version of this
  paper.}\BibitemShut {Stop}%
\bibitem [{\citenamefont {Varela}(2013)}]{GRQUICK}%
  \BibitemOpen
  \bibfield  {author} {\bibinfo {author} {\bibfnamefont {J.}~\bibnamefont
  {Varela}},\ }\href {https://library.wolfram.com/infocenter/ID/8329/}
  {\bibinfo {title} {{GRQUICK}}} (\bibinfo {year} {2013}),\ \bibinfo {note}
  {{Wolfram Library Archive, MathSource 8329}}\BibitemShut {NoStop}%
\bibitem [{Note2()}]{Note2}%
  \BibitemOpen
  \bibinfo {note} {The computations are carried out in the Mathematica notebook
  \protect \texttt {RG\protect \_UHCS\protect \_gravity.nb}, which is included
  as an ancillary file with the arXiv version of this paper.}\BibitemShut
  {Stop}%
\bibitem [{Note3()}]{Note3}%
  \BibitemOpen
  \bibinfo {note} {Had we instead set $\mu =\protect \sqrt {\omega }$ in
  $C_1(\mu )$, we would have derived $z=\protect \frac 2{1+3/(4|\kappa
  |)}={2-\protect \frac {3}{2|\kappa |}+O (|\kappa |^{-2})}$, which differs
  from the $\mu =|\protect \vec {k}|$ result only by $O(|\kappa |^{-2})$,
  beyond the accuracy of a one-loop calculation.}\BibitemShut {Stop}%
\bibitem [{Note4()}]{Note4}%
  \BibitemOpen
  \bibinfo {note} {Different values of $C_1$ at the fixed point are equivalent
  through time rescaling $t\to \lambda t$. At the one-loop level, the fixed
  point corresponds to a vanishing coefficient $1/C_2^2$ of the Calabi energy
  term, or equivalently, to $\protect \widetilde {\kappa }^2=\infty $. At the
  two- and higher-loop levels, it is possible that the fixed point moves to a
  nonzero $1/C_2^2$, i.e., a finite $\protect \widetilde {\kappa }^2$. As long
  as $\protect \widetilde {\kappa }^2 \gg 1$ at the fixed point (which should
  be the case for $|\kappa |\gg 1$), our prediction for the dynamical critical
  exponent, obtained by demanding the vanishing of the beta function for $C_1$,
  still stands.}\BibitemShut {Stop}%
\end{thebibliography}%

\end{document}